 \documentclass{article}

\usepackage{epsfig} 
\usepackage{amsfonts}
\usepackage{graphicx}

 \usepackage{amssymb}

\date{\today}

\newcommand{\insertplot}[5]{\begin{figure}
 \hfill\hbox to 0.05in{\vbox to #5in{\vfill
 \inputplot{#1}{#4}{#5}}\hfill}
 \hfill\vspace{-.1in}
 \caption{#2}\label{#3}
 \end{figure}}
 \newcommand{\inputplot}[3]{
 \special{ps: plotfile #1}
}
\newcounter{fig}   
 
\newcommand{\vphi}{\varphi}

\newcommand{\ee}{\end{equation}}
\newcommand{\eea}{\end{eqnarray}}
\newcommand{\be}{\begin{equation}}
\newcommand{\bea}{\begin{eqnarray}}

\begin{document}

\title{ 
Myers-Perry black holes with scalar hair and a mass gap
} 
  
\author{
{\large Yves Brihaye}$^{1}$,
{\large Carlos Herdeiro}$^{2}$
and
{\large Eugen Radu}$^{2}$
\\ 
$^{1}${\small Physique-Math\'ematique, Universite de
Mons-Hainaut, Mons, Belgium}
\\
$^{2}${\small Departamento de F\'\i sica da Universidade de Aveiro and I3N, 
   Campus de Santiago, 3810-183 Aveiro, Portugal}
}

\maketitle

\begin{abstract} 
We construct a family of asymptotically flat, rotating black holes with scalar hair and a regular horizon, within five dimensional Einstein's gravity minimally coupled to a complex, massive scalar field doublet. 
These solutions are supported by rotation and have no static limit. They are described by their mass $M$, two equal angular momenta $J_1=J_2\equiv J$ and a conserved Noether charge $Q$, measuring the scalar hair. For vanishing horizon size the solutions reduce to five dimensional boson stars. In the limit of vanishing Noether charge \textit{density}, the scalar field becomes \textit{point-wise} arbitrarily small and the geometry becomes,
\textit{locally}, arbitrarily close to that of a specific set of Myers-Perry black holes (MPBHs); but there remains a global difference with respect to the latter, manifest in a finite mass gap. Thus, the scalar hair never becomes a linear perturbation of MPBHs. This is a qualitative difference when compared to Kerr black holes with scalar hair~\cite{Herdeiro:2014goa}. Whereas the existence of the latter can be anticipated in linear theory, from the existence of scalar bound states on the Kerr geometry (\textit{i.e.} scalar clouds), the hair of these MPBHs is intrinsically non-linear. 
\end{abstract}

\section{Introduction and motivation}
Myers-Perry black holes (MPBHs)~\cite{Myers:1986un} have played an important role in exploring higher dimensional gravity. They represent the natural generalization of the Kerr solution~\cite{Kerr:1963ud} to higher dimensions. As such, MPBHs are a laboratory for testing how the dimensionality of spacetime affects Einstein's gravity and black hole (BH) physics at a fully non-linear level, and thus how special the four dimensional ($d=4$) Kerr solution is.

The different nature of the gravitational interaction in higher dimensions indeed endows MPBHs with a number of qualitatively distinct properties, when compared the Kerr solution.  One such distinction can already be seen in the zero angular momentum limit, \textit{i.e.} at the level of the Tangherlini solutions~\cite{Tangherlini:1963bw}. Whereas the $d=4$ Schwarzschild BH admits stable bound orbits for classical test particles, the same does not hold for the $d>4$ Tangherlini solutions~\cite{Tangherlini:1963bw}. This distinction is a direct consequence of the faster fall-off -- and hence shorter range -- of the gravitational interaction in higher dimensions. A similar distinction is also present at the level of gravitationally trapped test \textit{fields}, rather than particles. But in this case, the distinction becomes clearer considering MPBHs \textit{vs} Kerr BHs, rather than their static limits. For concreteness, we consider a $d=5$ MPBH with two equal angular momenta, a case for which there is symmetry enhancement. Also for concreteness, we take the field to be a massive, complex scalar field doublet, $\Psi$. But the arguments below apply more generally. 

Gravitationally trapped test field modes, \textit{i.e.} modes exhibiting an exponential fall-off at spatial infinity, have complex frequencies, and are called \textit{quasi-bound states}. The imaginary part of the frequency measures their (inverse) life-time and signals that in a BH background, gravitationally trapped fields are, in general, not infinitely long-lived, since they are infalling into the BH. In the four dimensional Kerr background, however, it is possible for a gravitationally trapped mode to become arbitrarily long-lived~\cite{Damour:1976kh,Zouros:1979iw,Detweiler:1980uk}, and even infinitely long-lived~\cite{Hod:2012px,Hod:2013zza,Herdeiro:2014goa}. If the mode is of the form $\Psi\sim e^{-iw t}e^{im\varphi}f(r,\theta)$,  in standard Boyer-Lindquist coordinates, the imaginary part of the frequency vanishes   when the real part obeys $w=m\Omega_H$, where $\Omega_H$ is the angular velocity of the BH horizon. Such modes, being infinitely long-lived, are true bound states, and have been called \textit{scalar clouds}. Scalar clouds occur at the threshold of the superradiant instability of Kerr BHs, which is an amplification of gravitationally trapped modes that occurs for $w<m\Omega_H$~\cite{Press:1972zz} (see~\cite{Cardoso:2013krh} for a review); thus they are boundary states, between gravitationally trapped decaying modes ($w>m\Omega_H$) and amplified modes. By contrast, for test massive scalar fields on the background of MPBHs, one cannot find gravitationally trapped modes in the superradiant regime~\cite{Cardoso:2005vk,Kunduri:2006qa}.  
Consequently, one cannot find scalar clouds around asymptotically flat MPBHs, a qualitative difference with the $d=4$ Kerr case.

The existence of scalar clouds around Kerr BHs, at the threshold of the superradiant instability, was taken in~\cite{Herdeiro:2014goa} as a smoking gun for the existence of Kerr BHs with scalar hair, as fully non-linear solutions of the Einstein-Klein-Gordon (EKG) system. These BHs were indeed constructed in~\cite{Herdeiro:2014goa}.  They form a five parameter family, classified by three continuous and two discrete parameters. The continuous parameters are the ADM mass $M$, ADM angular momentum $J$ and a scalar Noether charge $Q$, that measures the scalar hair. The two discrete parameters, labelled $m,n$, describe the azimuthal harmonic index and the number of nodes of the scalar field, respectively.  When $Q\rightarrow 0$, the solution reduces to a specific set of Kerr BHs, corresponding to the Kerr solutions which allow the existence of the corresponding scalar clouds. When the horizon radius shrinks to zero, the solutions reduce to $d=4$, asymptotically flat, rotating boson stars~\cite{Yoshida:1997qf,Kleihaus:2005me} (see also~\cite{Liebling:2012fv} for a recent review on boson stars). These Kerr BHs with scalar hair are therefore continuously connected to both Kerr BHs and boson stars, interpolating between these two types of solutions. They evade standard no-(scalar)-hair theorems due to the harmonic time dependence of the scalar field, which, however, disappears at the level of the scalar energy momentum tensor and also for the geometry. As such, even though the geometry is preserved by both a time-like (at infinity) and an azimuthal Killing vector field, the full solution is only preserved by a combination of these, yielding a helicoidal Killing vector field. The choice $w=m\Omega_H$, which in linear theory is a necessary condition to obtain scalar clouds, means -- for the general non-linear solution -- that the helicoidal Killing vector field is the null generator of the horizon, and thus the invariance of the scalar field along the orbits of this Killing vector field implies there is no scalar flux through the horizon.

The \textit{inexistence} of scalar clouds, for massive, test scalar fields on the background of asymptotically flat MPBHs\footnote{For asymptotically Anti-de Sitter ($AdS$) MPBHs, there are superradiant instabilities and indeed asymptotically $AdS_5$ MPBHs with scalar hair, and continuously connected to both $AdS$ boson stars and $AdS$ MPBHs, were reported in~\cite{Dias:2011at} as the first example of BHs with a single Killing vector field.}, implies that there are no MPBHs with scalar hair continuously connected to the vacuum solution. But, as we shall report in this paper, there are actually MPBH with scalar hair, albeit \textit{not continuously connected}, in terms of their global charges, with the vacuum MPBHs. 

We shall construct, by solving numerically the EKG equations in $d=5$, a three parameter family of asymptotically flat, regular (on and outside the horizon) BHs with scalar hair. The three continuous parameters are the ADM mass $M$, one ADM angular momentum $J$ parameter, which equals the two independent angular momentum parameters in $d=5$, and the Noether charge $Q$, which again measures the scalar field outside the horizon. Two discrete parameters, analogous to those mentioned above for Kerr BHs with scalar hair, could in principle be added as well. But here we shall focus on a single value of these parameters $(m=1,n=0)$ (\textit{c.f.} Sec.~\ref{sec_model}).  When the horizon size vanishes, the solutions reduce to a set of the asymptotically flat, $d=5$ rotating boson stars found in~\cite{Hartmann:2010pm}, in complete analogy to the horizonless limit of the Kerr BHs with scalar hair. Unlike the latter, however, the $d=5$ solutions we shall exhibit do not admit a limit of vanishing Noether charge. But they do admit a limit of vanishing Noether charge \textit{density}. In that limit the scalar field becomes, point-wise, arbitrarily small, and the geometry becomes, locally, arbitrarily close to that of a set of vacuum MPBHs. But the global charges -- mass, angular momentum and Noether charge -- have a finite gap, when compared to the global charges of the MPBHs that are locally approached. For these reasons we dub these solutions as \textit{MPBHs with scalar hair and a mass gap}. 

The behaviour we have just described is distinct from that observed for Kerr BHs with scalar hair, and it makes the existence of hairy MPBHs consistent with the absence of (test) scalar clouds around vacuum MPBHs. This behaviour is anchored on the shorter range of the gravitational interaction in $d\ge 5$. The higher dimensional shorter range gravity may still produce gravitational trapping; but this requires non-linear effects.

\section{The model}
\label{sec_model}
 
We consider $d=5$ Einstein's gravity minimally coupled to a massive complex scalar field doublet, $\Psi$. The system is described by the action  
\begin{equation}
\label{action}
S=\int  d^5x \sqrt{-g}\left[ \frac{1}{16\pi G}R
   -\frac{1}{2} g^{ab}\left( \Psi_{, \, a}^* \Psi_{, \, b} + \Psi _
{, \, b}^* \Psi _{, \, a} \right) 
-U(|\Psi| )
 \right] ,
\end{equation}
where $G$ is Newton's constant (that will be set to unity)
and $U(|\Psi|^2)$ is the scalar field potential.
Variation of this action with respect to the metric and scalar
field gives the EKG equations:
\begin{equation}
\label{EKG-eqs}
 R_{ab}-\frac{1}{2}g_{ab}R=8 \pi ~T_{ab},
~~\left(\Box -\frac{d U }{d  |\Psi|^2}\right)\Psi=0~,
\end{equation}  
where 
\begin{equation}
\label{Tab}
T_{ab}= 
 \Psi_{ , a}^*\Psi_{,b}
+\Psi_{,b}^*\Psi_{,a} 
-g_{ab}  \left[ \frac{1}{2} g^{cd} 
 \left( \Psi_{,c}^*\Psi_{,d}+
\Psi_{,d}^*\Psi_{,c} )+U(|\Psi| 
\right)
\right]
\end{equation}
is the
 stress-energy tensor  of the scalar field.
 

The solutions we report in this work are described by a metric ansatz\footnote{
An equivalent form of the metric (related to the one used in 
\cite{Dias:2011at})
reads
\begin{eqnarray}
\nonumber
ds^2&=&e^{2F_1(r)} \frac{dr^2}{N(r) } 
+e^{2F_1(r)}r^2\frac{1}{4} 
\left(
\sigma_1^2+\sigma_2^2
\right)
+e^{2F_2(r)}r^2\frac{1}{4} (\sigma_3-2W(r)dt)^2
-e^{2F_0} N(r) dt^2,
\end{eqnarray}
with the left-invariant 
1-forms $\sigma_i$ on $S^3$, 
$\sigma_1=\cos  \psi d  \Theta+\sin \psi \sin  \Theta d \varphi$,
$\sigma_2=-\sin  \psi d  \Theta+\cos \psi \sin   \Theta d  \varphi$,
$\sigma_3=d \psi  + \cos  \Theta d  \varphi$  
 and $\Theta=2\theta, \varphi=\varphi_2-\varphi_1, \psi=\varphi_1+\varphi_2)$ the usual Euler angles.
 The corresponding scalar field ansatz is
\begin{equation}
 \nonumber
 \Psi = \phi(r) e^{i\left(\frac{\psi}{2}-w t\right)} 
 \left( \begin{array}{c} 
   \sin\frac{\Theta}{2}  e^{-i \frac{\vphi}{2}} \\    \sin\frac{\Theta}{2}  e^{i \frac{\vphi}{2}} 
 \end{array} \right) .
\end{equation} 
} 
\begin{eqnarray}
\label{ansatz}
ds^2&=&e^{2F_1(r)}\left(\frac{dr^2}{N(r) }+r^2 d\theta^2\right)+e^{2F_2(r)}r^2 
\left[
\sin^2\theta (d\varphi_1-W(r) dt)^2
+
\cos^2\theta (d\varphi_2-W(r) dt)^2
\right] \nonumber
\\
&&
+(e^{2F_1(r)}-e^{2F_1(r)})\sin^2\theta \cos^2\theta (d\varphi_1-d\varphi_2)^2
-e^{2F_0} N(r) dt^2,
\end{eqnarray}
with 
\begin{eqnarray}
\nonumber
N(r)=1-\frac{r_H^2}{r^2},
\end{eqnarray}
where $\theta  \in [0,\pi/2]$, $(\varphi_1,\varphi_2) \in [0,2\pi]$, 
and $r$ and $t$ denote the radial and time coordinate, respectively.\footnote{
The Myers-Perry vacuum solution can be expressed as the line element (\ref{ansatz}) with
\begin{eqnarray}
\nonumber
e^{-2F_0(r)}=e^{4F_1(r)}+\frac{M^2-\frac{r_H^4}{4}}{r^4} , \ \ \  e^{2F_1(r)}=1+\frac{M-\frac{r_H^2}{r^2}}{r^2}, \ \ \  F_2(r)=-(F_0(r)+F_1(r)), \ \ \ 
W(r)= \frac{\sqrt{2M}}{r^4}\sqrt{M^2-\frac{r_H^4}{4}}
e^{-2F_1(r)}.
\end{eqnarray}
} 
The scalar field ansatz is a complex doublet first used in~\cite{Hartmann:2010pm} to obtain rotating bosons stars 
\begin{equation}
 \Psi = \phi(r) e^{-i w  t} 
 \left( \begin{array}{c} 
   \sin\theta ~ e^{i \vphi_1} \\ \cos\theta ~ e^{i \vphi_2} 
 \end{array} \right) 
 , \label{phi} 
\end{equation} 
where $\phi$ is a real function  and $w>0$ is the frequency. Observe that this ansatz takes the azimuthal harmonic index on both planes to be $m=1$, as mentioned in the Introduction. Also, in accordance to the discussion therein, we shall look for solutions such that $\phi(r)$ is nodeless, \textit{i.e.} $n=0$.
%
For such an ansatz with two equal angular momenta,
the spacetime isometry group is enhanced from $\mathbb{R}_t \times U(1)^{2}$
to $\mathbb{R}_t \times U(2)$, where $\mathbb{R}_t$ denotes the time translation.
But observe this is not a symmetry of the scalar field. The full 
ansatz is preserved only by a helicoidal Killing vector field, of the form 
$\chi =\partial_t+w( \partial_{\varphi_1}+ \partial_{\varphi_2})$.

The $AdS_5$ MPBH solutions in \cite{Dias:2011at}
have been found for a rather similar metric ansatz
and a vanishing scalar potential,
$U(|\Psi| )=0$; in this case spacetime asymptotics supply the required
confining mechanism so that scalar clouds can be found at linear level. For asymptotically  Minkowski spacetimes, a mass term for the scalar field is the simplest way to obtain gravitationally trapped states. Thus we take 
\begin{eqnarray}
\label{potential}
U(|\Psi| )= \mu^2 \Psi^*\Psi,
\end{eqnarray}
where $\mu$ is the scalar field mass.
We remark that the asymptotically flat boson stars in 
\cite{Hartmann:2010pm}
have been constructed for a more general potential, which 
is chosen such that non-topological soliton 
solutions -- dubbed \textit{Q-balls} -- 
exist in the absence of gravity (also, the work \cite{Hartmann:2010pm} uses a 
different scaling convention). 
Even though similar solutions to the ones we shall describe could be constructed for this more general potential, to simplify the picture, we shall restrict our study here 
  to the case of the potential (\ref{potential}).

\section{The solutions}

\subsection{Boundary and physical conditions}

We seek asymptotically flat solutions. As such we impose that $F_i=W=\phi=0$, as $r\rightarrow \infty$.
The ADM mass $M$ and angular momenta $J_1=J_2=J$, are read from the asymptotic behaviour of the metric functions,
\begin{eqnarray}
\label{asym}
g_{tt} =-1+\frac{8 M}{3\pi r^2}+\dots,
~~g_{\varphi_1 t}=-\frac{4 J}{\pi r^2}\sin^2\theta+\dots,
~~g_{\varphi_2 t}=-\frac{4 J}{\pi r^2}\cos^2\theta+\dots~.
\end{eqnarray}
For the scalar field, the asymptotic behaviour is 
\begin{equation}
\phi=c_1\frac{e^{-\sqrt{\mu^2-w^2}r}}{r^{3/2}}+\dots ,
\label{scadec}
\end{equation}
with $c_1$ a constant. As expected, the condition $w <\mu $ must hold, for gravitationally trapped modes.

We also seek solutions that are regular on the horizon. In our coordinates, the event horizon resides at a surface of constant  radial variable $r=r_H>0$. To simplify the numerical treatment of the problem, we introduce a new radial coordinate $x=\sqrt{r^2-r_H^2}$, such that the horizon is located at $x=0$.
Then a power series expansion near the horizon yields, imposing regularity,  
$F_i=F_i^{(0)}+{\cal O}(x^2)$,
$W=\Omega_H +{\cal O}(x^2)$,
and $\phi=\phi_0+{\cal O}(x^2)$,
with 
\begin{eqnarray}
\label{cond}
w=\Omega_H,
\end{eqnarray}
where $\Omega_H>0$ is the horizon angular velocity.
Thus the boundary conditions at the horizon are  
${dF_i}/{dx} = {d\phi}/{dx} =  0$ and
$W=w$. Observe that with the choice (\ref{cond}) the Killing vector field $\chi$ becomes the null horizon generator.  Then, the preservation of the scalar field along the orbits of this field, $\chi^{\mu}\partial_\mu \Psi=0$, implies the absence of scalar flux through the horizon.
 
The presence of a horizon introduces a temperature $T_H$ and an entropy $S=A_H/4$, where
\begin{eqnarray}
\label{THAH}
T_H=\frac{e^{F_0^{(0)} -F_1^{(0)}}}{2\pi r_H}, \qquad A_H=2\pi^2 r_H^3   e^{2F_1^{(0)}+F_2^{(0)} }.
\end{eqnarray}
Also, the  Lagrangian of the scalar field  has a global $U(1)$ symmetry which introduces a
conserved current $j^a=-i (\Psi^* \partial^a \Psi-\Psi \partial^a \Psi^*)$,
with $j^a_{;a}=0$.
Thus the solutions carry also a conserved Noether charge, obtained by integrating the Noether charge density, $j^t$, on a spacelike slice $\Sigma$,  
\begin{eqnarray}
\label{Q}
Q=\int_{\Sigma}dr d\theta d\varphi_1d\varphi_2  ~j^t \sqrt{-g}=
4\pi^2 \int_{r_H}^\infty dr~r^3  \phi^2(r)e^{-F_0(r)+3F_1(r)+F_2(r)}   \frac{(w-W(r))}{N(r)}.
\end{eqnarray}
The temperature, entropy and the global charges
are related through the Smarr mass formula 
\begin{eqnarray}
\label{smarr}
\frac{2}{3}M=T_H S +\Omega_H (2J-Q)+ \frac{2}{3}M^{(\Psi)},
\end{eqnarray}
 where
 \begin{eqnarray}
M^{(\Psi)}=-\frac{3}{2}\int_{\Sigma}  \sqrt{-g} dr d\theta d\varphi_1d\varphi_2 \left( T_t^t-\frac{1}{3}T_a^a \right),
\end{eqnarray}
is the energy stored in the scalar field
outside the BH horizon, and also by the first law of BH thermodynamics
\begin{eqnarray}
\label{fl}
dM=T_H dS +2\Omega_H dJ .
\end{eqnarray}

\subsection{Remarks on numerics}
The numerical  approach we have employed here is similar to that used 
in constructing $d=5$ BH solutions with equal-magnitude angular momenta in
Einstein-Maxwell-Chern-Simons theory 
\cite{Blazquez-Salcedo:2013muz}
or in Einstein-Gauss-Bonnet theory
\cite{Brihaye:2013vsa}.
As usual, dimensionless variables and global quantities are introduced by using
natural units set by $\mu$ (we recall $G=1$), $e.g.$ $r\to r/\mu$,
 $\phi \to \phi  /\sqrt{8\pi}$ and $w \to w/\mu$.
Then,  the numerical treatment of the model relies on only two input parameters:   
the horizon radius $r_H$ and the field frequency $w$.

The approach used in the asymptotically $AdS$ case \cite{Dias:2011at} (see also \cite{Stotyn:2011ns})
to construct a closed form  perturbative expression for the solutions
does not apply in the asymptotically flat case.\footnote{Approximate expressions for the solutions
can be written for $r\to r_H$ and $r\to\infty$.
However, since one cannot match the parameters in these asymptotic expansions,
the corresponding expressions are not really useful.} As such, we have to rely on numerical methods to obtain asymptotically flat hairy MPBHs. Thus, to guarantee the accuracy of the solutions reported here, these have been obtained using two different solvers and numerical methods.

The system of five non-linear coupled differential equations
for the functions  $F_i,W$ and $\phi$, subject to the  boundary conditions described above,
was solved first by using the software package COLSYS developed
by Ascher, Christiansen and Russell \cite{COLSYS}.
This solver employs a collocation
method for boundary-value ordinary differential equations and a damped Newton method
of quasi-linearization. At each iteration step a linearized problem is solved by using a spline
collocation at Gaussian points.
Typical meshes use around 400 points in the interval $0\leq x<x_{max}$,
with $x_{max}$ around $10^{4}$ (we have verified
that the solutions are independent on the choice of $x_{max}$).
Also, the  mesh is non-equidistant, being denser close to the horizon.

As a crosscheck of the results, a large part of the solutions were also derived
 with the program FIDISOL,
which uses a  Newton-Raphson method.
This software provides also an error estimate for each unknown function
(see \cite{schoen}  
for details and examples for the numerical procedure).
In this case we have introduced a compactified coordinate 
$\bar{x}=x/(1+x)$, with $0\leq \bar{x} \leq 1$.
Typical grids used have  around $250$ points,
distributed equidistantly over the 
 full integration region. 
 
We have found a very good agreement between the solutions constructed with these two different methods. 
In both cases,
apart from convergence tests, the Smarr relation (\ref{smarr})
and the first law (\ref{fl})
have been used to test 
the accuracy of the results.
Based on that, we estimate a typical relative error $<10^{-4}$
for the solutions reported herein.

\subsection{The five dimensional rotating asymptotically flat boson stars}
\label{sec_bs}
Before discussing the MPBHs with scalar hair and a mass gap, it is useful
to review the basic properties of the solitonic limit of the solutions.
In this case $r_H=0$ and the horizon is replaced with a regular origin, 
where $d F_i/dr =W=\phi=0$. 
As shown in \cite{Hartmann:2010pm},  the Noether charge and the angular momenta of these boson stars are not independent quantities; they are simply related by 
\begin{eqnarray}
Q=2J,
\end{eqnarray}
 while the Smarr relation and the first law read
\begin{eqnarray}
 M=M^{(\psi)},~~dM=2 w dJ.
\end{eqnarray}

\begin{figure}[h!]
\centering
\includegraphics[height=2.48in]{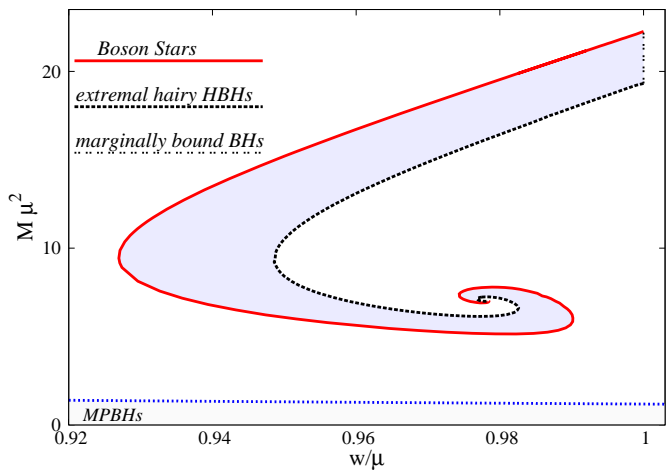}\\
\caption{Line of existence of boson stars (red solid) and domain of existence of hairy MPBHs (shaded area)  in $(M,w)$-space. Vacuum MPBHs exist below the almost horizontal (blue dotted) line at the bottom of the diagram.  
} 
\label{Mw}
\end{figure}

Taking $w$ as a control parameter, the numerical results show that boson stars exist for a limited range of frequencies,
$w_{min}  <w< \mu$,
with $w_{min}/\mu\simeq 0.927$ - Figure \ref{Mw} (red solid line). 
The most striking
property of the $d=5$ boson stars is that these configurations do not connect continuously to Minkowski spacetime.
In contrast to the $d=5$ $AdS$ case 
\cite{Dias:2011at},
or with the $d=4$ asymptotically flat case 
\cite{Yoshida:1997qf,Kleihaus:2005me},
the $d=5$ asymptotically flat boson stars do not trivialise as the maximal frequency is reached, 
$w\to \mu$.
Indeed, as discussed already in  \cite{Hartmann:2010pm},
in this limit the scalar field spreads and tends to zero \textit{point-wise}, 
while the geometry becomes arbitrarily close to that of flat spacetime. But in this limit, the global charges of the solutions remain finite and nonzero. 
Thus a mass, Noether charge and angular momentum gap is found between the $\phi=0$ vacuum Minkowski ground state and the limiting configurations with a frequency $w$ arbitrarily close to $\mu$.

An analytical understanding of this feature was given in  Ref. \cite{Hartmann:2010pm},
and relies basically on the special scaling properties of the EKG system 
in $d=5$ spacetime dimensions.
Following the $d=4$ analysis in \cite{Friedberg:1986tp},
one notices that as $w\to \mu$,
the solution scales as 
\begin{eqnarray}
\label{scaling}
F_i=F_{i0}+\xi^2  F_{i2}+\dots ,~~
W=W_{0}+\xi ^4 \tilde W+\dots,~~
\phi=\xi^2 \tilde \phi+\dots,~~
r=\tilde r/\xi,~~{\rm with}~~\mu^2=w^2+\xi^2 \hat w_c^2
\end{eqnarray} 
where $\xi$ a small parameter and $\hat w_c$ a constant.
Also $F_{i0}=W_{0}=0$, for a Minkowski background.
Substituting in the EKG equations yields  to first order in $\xi^2$  
a system of three coupled equations for 
$F_{02}$,
$\tilde W$
and 
$\tilde \phi$ (with $F_{12}=F_{22}=-F_{02}/2$),
which are solved in Ref.
\cite{Hartmann:2010pm}.
The profile of $\tilde \phi$ is then used to evaluate  
the Noether charge of the limiting solutions,
$Q=4 \pi^2 \mu \int_0^{\infty}  dr~ r^3 \phi^2$, which is nonzero
despite the fact that $\phi$ is infinitesimally small. This can be understood by noticing that $Q$ is not affected by the scaling (\ref{scaling}). 
Thus, whereas the Noether charge \textit{density} becomes infinitesimally small as $w\rightarrow \mu$,
 the Noether charge remains finite and non-zero. 

A similar relation can be written for the mass  $M=M^{(\psi)}$ of the limiting boson stars.
Then,
as proven in \cite{Hartmann:2010pm},
 the solutions
exhibit an universal behaviour, independent on the details of the scalar field potential,  
with
\begin{eqnarray}
\label{Mmax}
M_{max}=\mu Q_{max},
\end{eqnarray}
 as $w\to \mu$.
 For the scaling convention used in this work, one finds
 $M_{max}\mu^2 \simeq 22.11$.

\begin{figure}[h!]
\centering
\includegraphics[height=2.48in]{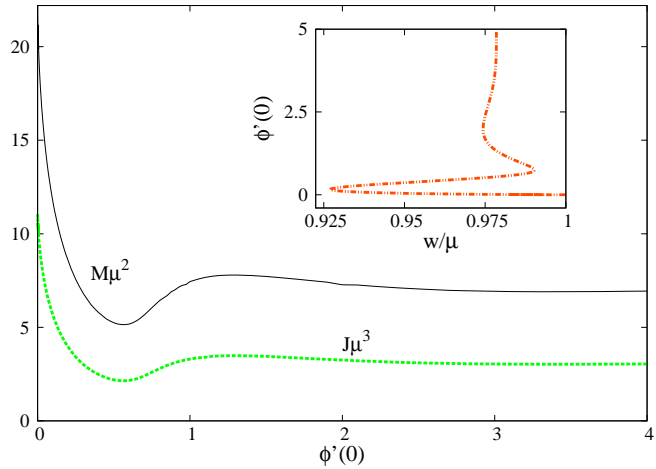}\\
\caption{The mass $M$ and angular momentul $J$ are shown  $vs.$  the derivative of
the scalar field at the origin $\phi'(0)$
for $d=5$ asympotically flat rotating boson stars. (Inset) The relation between $\phi'(0)$ and the control parameter $w$.} 
\label{phiprime}
\end{figure}

As can be seen in Figure \ref{Mw} (red solid curve), 
the boson star mass decreases as $w$ is decreased from the maximal value $\mu$.
Then, after approaching $w_{min}$, a
backbending is observed in the $M(w)$ diagram. Further following the curve, there is an inspiralling behaviour, towards a limiting configuration
at the center of the spiral, which occurs for a frequency $w_{cr}/\mu \simeq 0.978$.
The central inspiralling behaviour is quite generic for boson stars\footnote{An exceptional case for which this  diagram changes is provided by solutions of the $d=5$ Einstein-Gauss-Bonnet model
 \cite{Brihaye:2013zha,Henderson:2014dwa}.}, 
being found also in $d=4$ Einstein gravity and in $d=5$
solutions with AdS asymptotics \cite{Dias:2011at}.
A similar diagram is found for $J(w)$; 
thus it is clear that the boson stars do not possess a slowly rotating limit.

Instead of using $w$, the global charges of these boson star solutions may be given in terms of $\phi'(0)$,
which is proportional to the scalar energy density at the centre of the boson star. The curves for the mass and angular momentum in terms of $\phi'(0)$, as well as the relation between $\phi'(0)$ and $w$, are plotted in Figure \ref{phiprime}. One can see that $\phi'(0)$ diverges as $w\to w_{cr}$, 
while both $M$ and $J$ are finite in that limit.

As a manifestation that although these non-linearly gravitationally trapped scalar fields exist, the trapping is weak in higher dimensions, one can show that the boson star mass is larger than the mass of $Q$
free bosons. Then the arguments in \cite{Hartmann:2010pm}
show that these solutions are classically unstable.

\subsection{Myers-Perry black holes with scalar hair and a mass gap}

The hairy MPBHs are constructed by using the numerical methods described above and the following general strategy. For any spinning boson star with frequency $w$, we have found that a small BH can be added at its centre. Moreover, for a fixed $w$, the boson star solution, which has  $r_H=0$,
provides a good initial profile for finding a hairy MPBHs with a small $r_H$.
By increasing $r_H$ from zero, we obtain rotating BH solutions with $\Omega_H$
fixed by (\ref{cond}). Proceeding in this way we have managed to scan the phase space of hairy MPBHs. 

\begin{figure}[h!]
\centering
\includegraphics[height=2.2in]{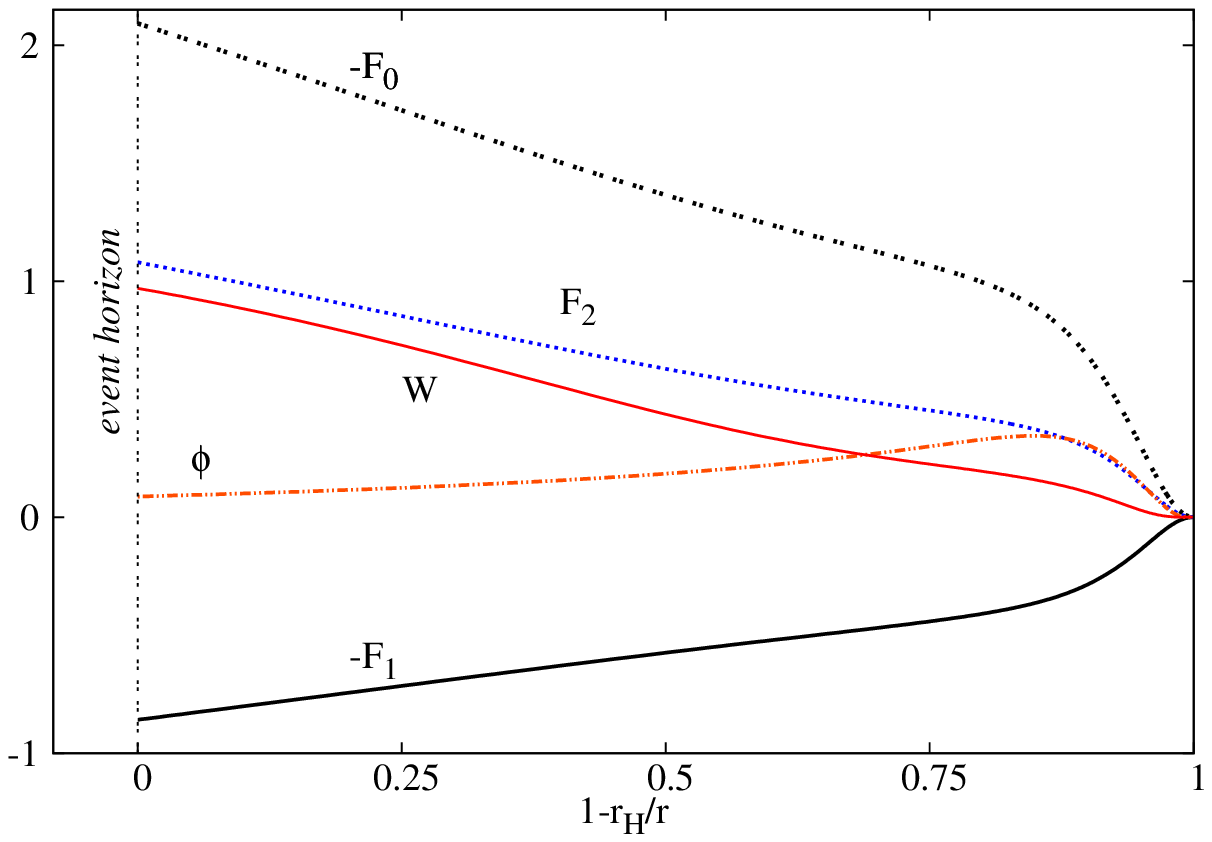}
\includegraphics[height=2.2in]{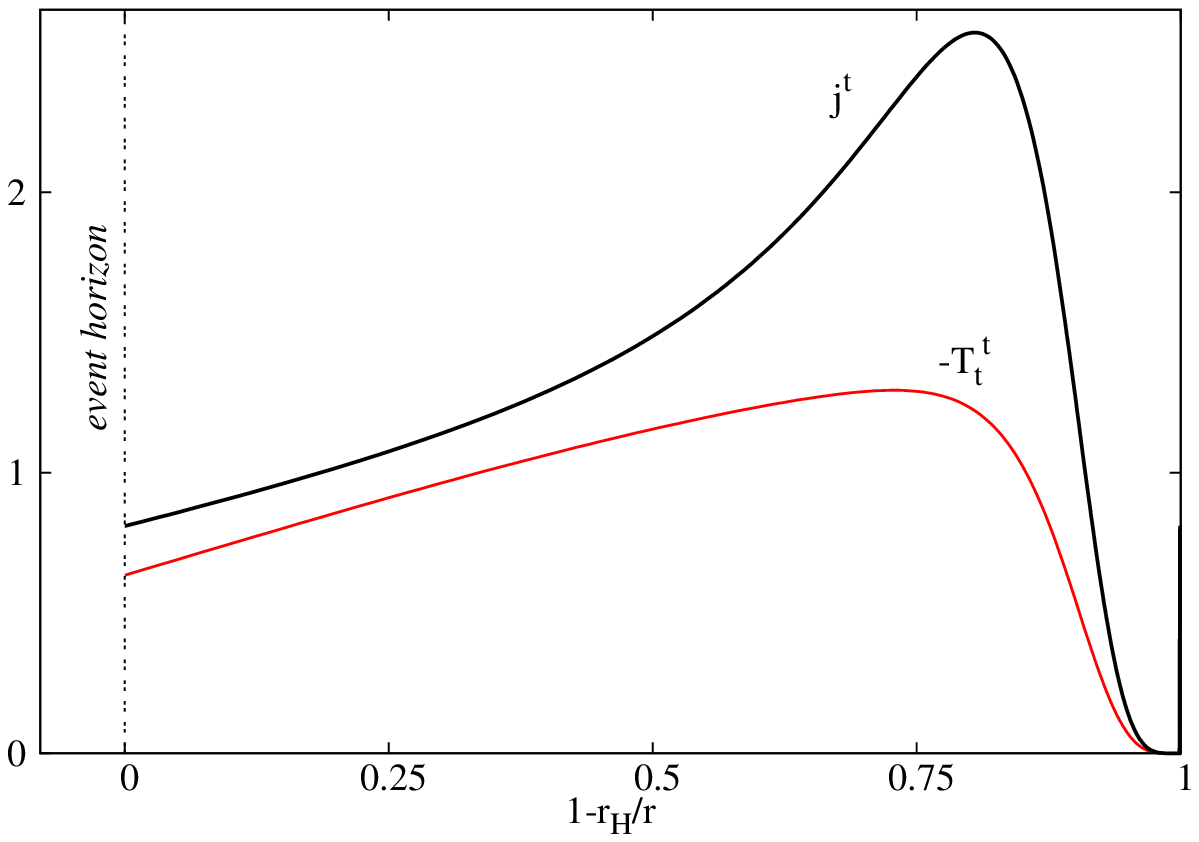}
\caption{The metric functions and the scalar field (left panel) are shown together with the energy and Noether charge
densities (right panel) for a 
 typical hairy MPBH solution.} 
\label{metricfunctions}
\end{figure}

For all the solutions we studied, the metric functions $F_i(r)$  
and $W(r)$ interpolate monotonically between the corresponding values 
at $r=r_H$ and the asymptotic value  at infinity, without presenting 
any local extrema. The scalar field, for the nodeless solutions discussed herein, is finite and non-zero at $r=r_H$, presents a local maximum outside the horizon and decays exponentially towards zero, asymptotically. 
As a typical example, in Figure \ref{metricfunctions} (left panel) the metric 
functions $F_i(r)$, $W(r)$ and the scalar function $\phi(r)$  are
shown for a solution with $r_H=0.125/ \mu$, $\Omega_H=0.97 \mu$, as functions of 
the scaled coordinate $1-r_H/r$.
The profiles of the corresponding energy density and Noether charge density are also shown (right panel).

In Figure \ref{Mw} (shaded region), we exhibit the domain of existence of hairy MPBHs, in a $M(w)$ diagram, based on several thousands of solution points. This domain is delimited by three curves: the already discussed boson star curve (red solid line), the curve of \textit{extremal} (\textit{i.e.} zero temperature) hairy MPBHs (black dashed line), and a vertical line segment with $w=\mu$ which correspond to the limiting configurations we dub \textit{marginally bound solutions} (black dotted line), for reasons we shall explain below. We remark that a similar diagram is found for $J(w)$ (with $J_{min}\mu^3\simeq 4.424$). 
Thus, we conclude that hairy MPBHs have a minimal mass and angular momentum. In particular they have no static limit, analogously to Kerr BHs with scalar hair \cite{Herdeiro:2014goa}.

The extremal solutions have finite
horizon size and global charges and  possess a regular horizon. The line describing these solutions has the same qualitative behaviour of the boson star line, described in Section \ref{sec_bs}: it starts from a non-zero mass at the maximal frequency, decreases until a minimal value of $w/\mu\simeq 0.947$, backbends and keeps decreasing, reaches a minimal value of the mass and then inspirals towards a central value where, we conjecture,\footnote{At the very centre of the inspiral numerics are increasingly challenging.} it meets the endpoint of the boson star spiral in a singular solution for $\Omega_H/\mu \simeq 0.978$.

\begin{figure}[h!]
\centering
\includegraphics[height=2.2in]{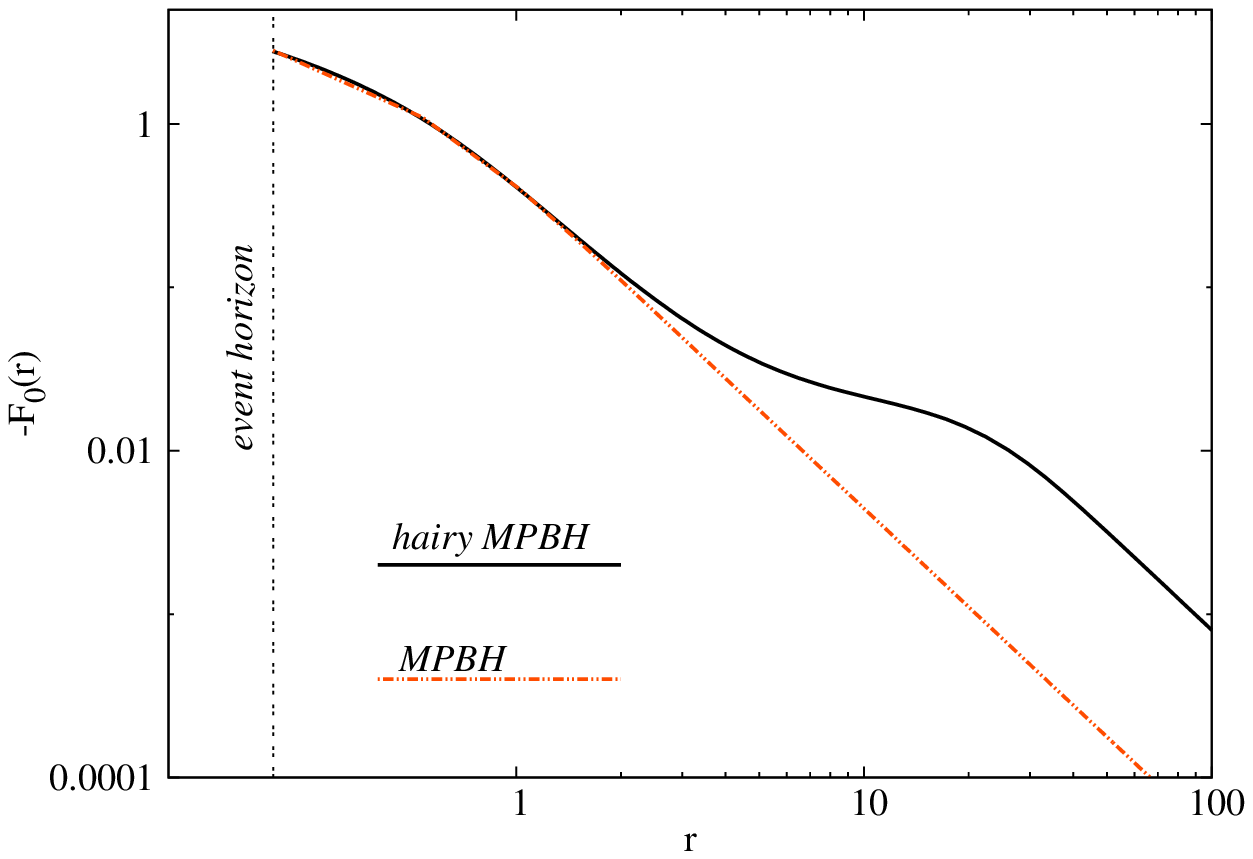}
\includegraphics[height=2.2in]{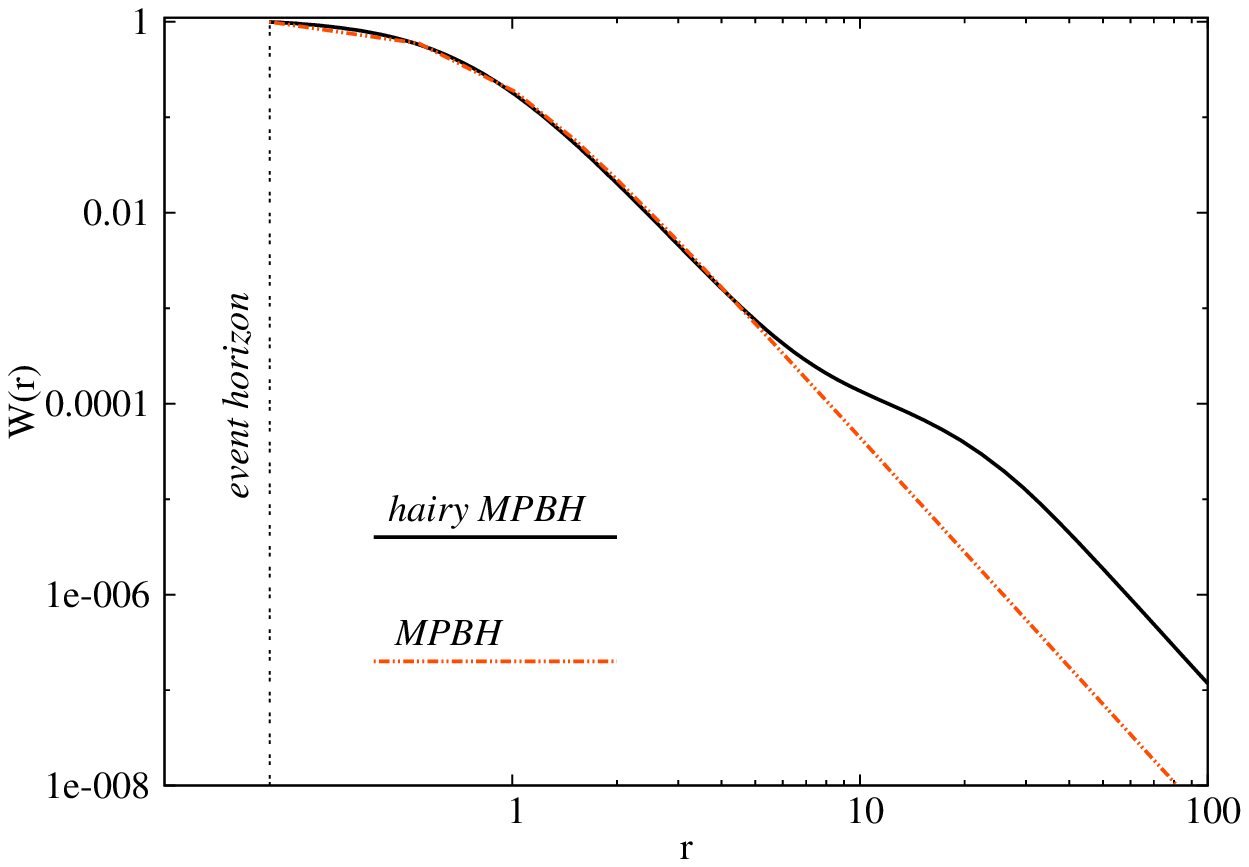}
\caption{ The metric functions $F_0$ and $W$ are shown for a vacuum MPBH and a hairy MPBH
close to the marginally bound set. Both solutions have $\Omega_H=0.995$ and $r_H=0.2$.}  
\label{metric_limit}
\end{figure}

\begin{figure}[h!]
\centering
\includegraphics[height=2.2in]{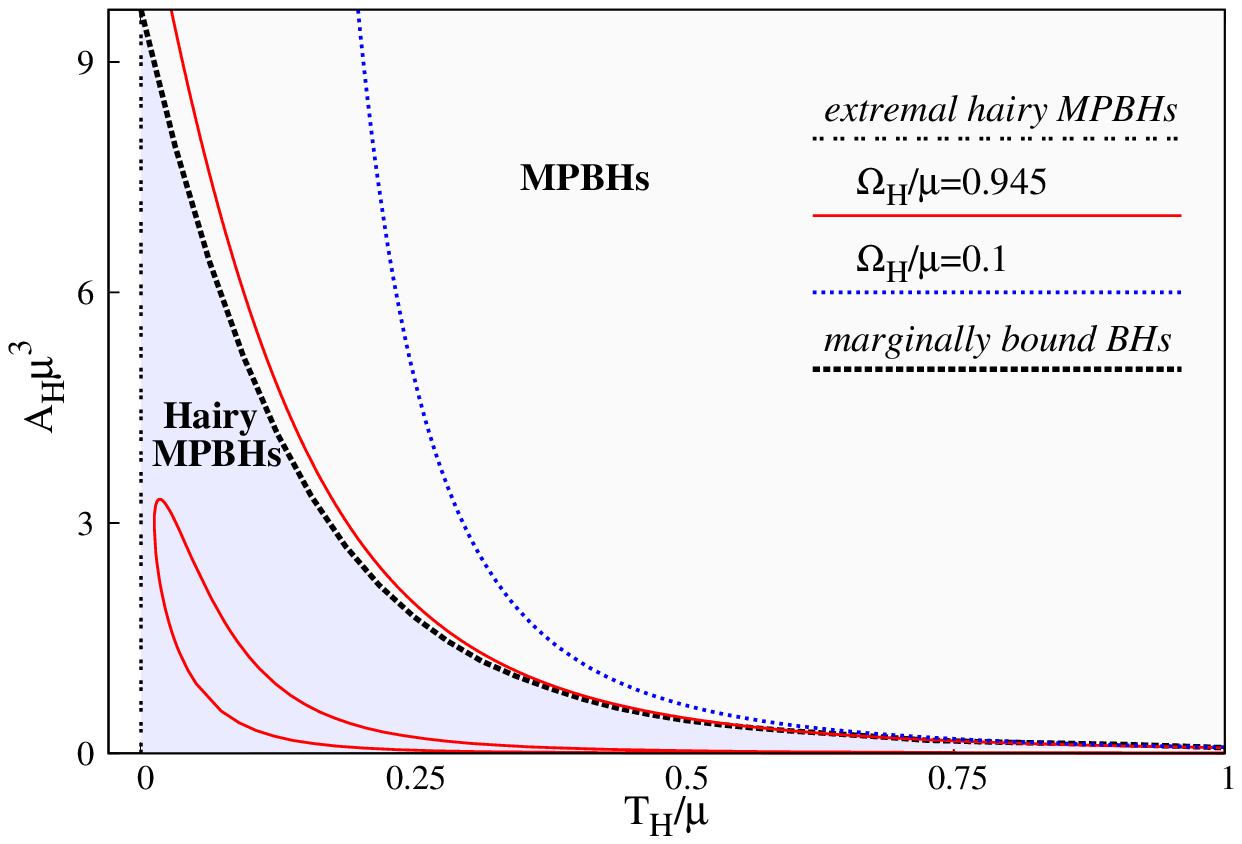}
\includegraphics[height=2.2in]{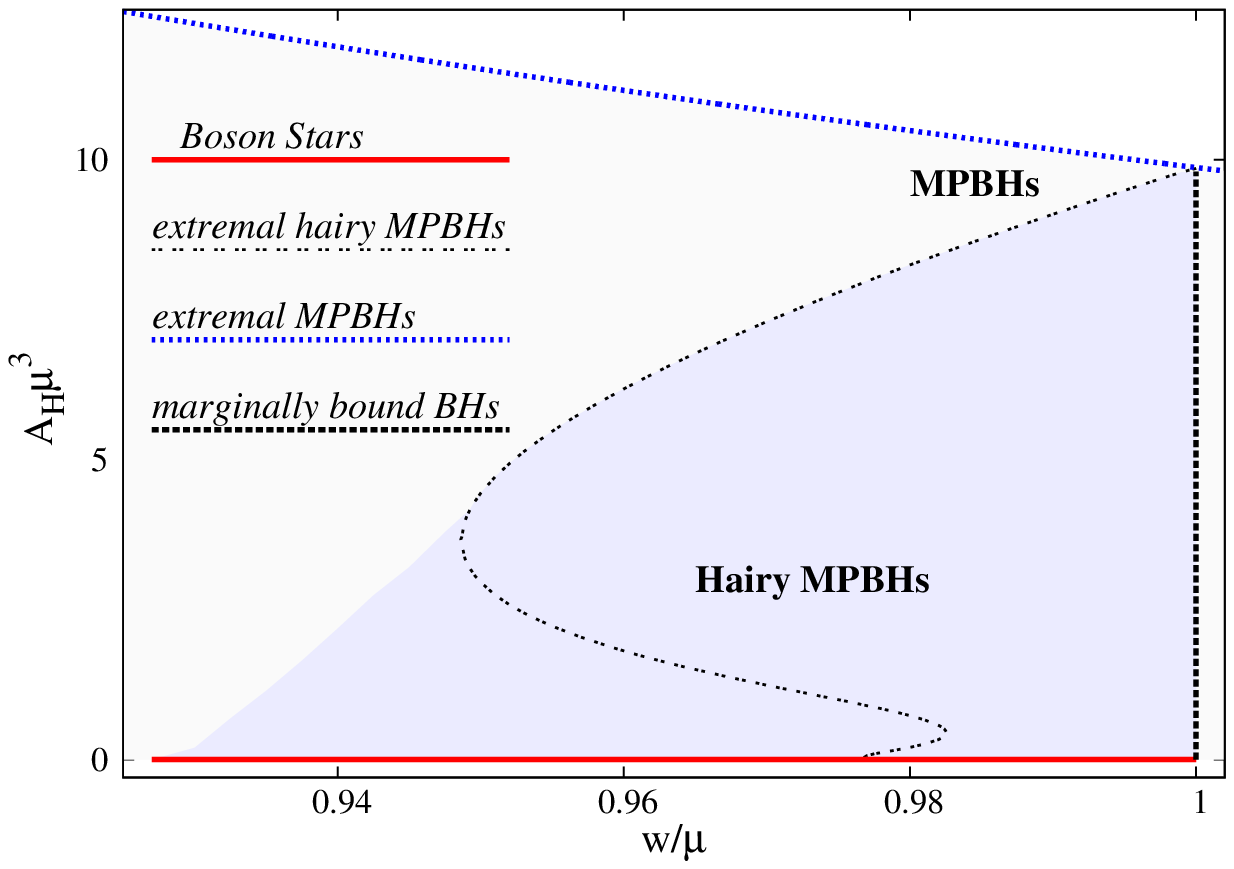}
\caption{(Left panel) Domain of existence (shaded area) of hairy MPBHs in a horizon area $A_H$ \textit{vs} temperature diagram. We have also plotted lines of constant horizon angular velocity $\Omega_H$. When $\Omega_H=\mu$ the line of vacuum MPBHs precisely coincides with that of the marginally bound hairy MPBH solutions. (Right panel) Domain of existence in a horizon area \textit{vs} frequency (equal to $\Omega_H$) diagram. Vacuum MPBHs exist below the blue dotted line. For $\Omega_H=\mu$, their domain of existence coincides precisely with the marginally bound hairy MPBHs.} 
\label{areatemperature}
\end{figure}

The marginally bound solutions are approached as $w\to \mu$. In this limit the scalar field ceases to be gravitationally trapped, \textit{c.f.} (\ref{scadec}), becoming only marginally bound.\footnote{This behaviour is reminiscent of that observed in \cite{Degollado:2013eqa,Sampaio:2014swa} for marginal \textit{test} scalar and Proca clouds around spherically symmetric charged BHs.}  These solutions provide a qualitative difference as compared to both Kerr BHs with scalar hair~\cite{Herdeiro:2014goa} and $AdS_5$ MPBHs with scalar hair~\cite{Dias:2011at}, since in both these cases the solutions trivialize as  $w\to \mu$.
The scalar field of these marginally bound solutions exhibits the same behaviour as found for boson stars.
Thus, for a  given  $r_H>0$, 
 the scalar field spreads and tends to zero   as $w$ increases towards $\mu$,
while the integral $\int_{r_H}^{\infty}  dr~ r^3 \phi^2$ is still nonzero
(in fact the value of this  integral is fixed by the parameter $r_H$ only).
At the same time, the geometry becomes that of a vacuum MPBH, instead that of that of flat Minkowski background as in the $r_H=0$ case. This can be seen in Figure \ref{metric_limit}, where we compare the behaviour of two (illustrative) metric functions for a hairy MPBH -- which is approaching the marginally bound solutions -- and a vacuum MPBH with the same horizon properties. The metric functions coincide near the horizon and a small difference occurs only asymptotically. As $w\rightarrow \mu$ the point at which the metric functions start to become different moves to a larger radial coordinate. Thus, for marginally bound solutions, the horizon properties, in particular the Hawking temperature and entropy, equal that of MPBHs with $\Omega_H=\mu$ - Figure \ref{areatemperature}.  As $r_H\to 0$, the limiting boson star is recovered, while  for 
$r_H\to r_H^{(max)}=\frac{\mu}{\sqrt{2}}$,
the geometry of the extremal vacuum MPBH is approached, albeit with different global charges.

To get an analytic understanding of this behaviour, we can, in principle, follow the analysis in \cite{Hartmann:2010pm}  
of the boson star limit, and consider the expansion (\ref{scaling}) of the solutions 
in the small parameter $\xi^2\sim \mu^2 - w^2$, together with the same scaling or $r$.
In this case, however, the functions $F_{i0},W_0$
do not vanish, being given by those of the corresponding MPBH with the same $r_H$ and $\Omega_H=\mu$.
Substituting in the EKG equations and taking the terms in $\xi^2$
leads to a system of five coupled equations to be solved with suitable
boundary conditions.
Since for $r_H\neq 0$  their study is rather cumbersome and we shall not pursue it here.
 
\begin{figure}[h!]
\centering
\includegraphics[height=2.48in]{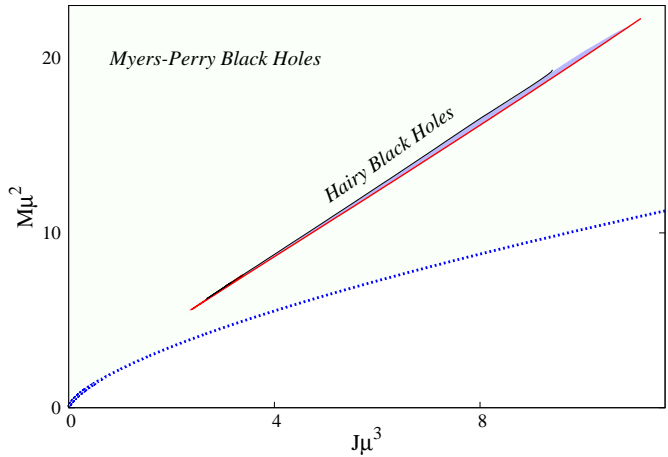}
\\
\caption{The domain of existence (very thin shaded area) of hairy MPBHs in the $(J,M)$ space. Vacuum MPBHs exist above the blue dotted line.} 
\label{jm}
\end{figure}

In Figure \ref{jm} we 
plot the phase space of the hairy MPBHs, \textit{i.e.} the domain of existence of these BHs  in the $(J,M)$-plane. As it can be observed they exist in the region where vacuum MPBHs exist as well. As such there is non-uniqueness, when only the ADM mass and angular momentum are specified, in analogy to the case of Kerr BHs with scalar hair. 
 
Other properties of the solutions are similar to 
the case of Kerr BHs with scalar hair.
For example, the scalar field is always spatially localized within the vicinity of the horizon. 
Geometrically, the horizon is a squashed sphere,
whose deformation can be parametrized by the ratio $F_2/F_1$.

\section{Further remarks}
The hairy MPBHs with a mass gap we have reported in this paper illustrate the qualitatively different behaviour of higher dimensional gravity. They could not have been anticipated from linear theory -- unlike Kerr BHs with scalar hair~\cite{Herdeiro:2014goa,Herdeiro:2014ima,Herdeiro:2014jaa} -- since a massive scalar field does not form test scalar clouds around vacuum MPBHs. The solutions herein rely on non-linear effects to achieve gravitational trapping. As such, they provide a new example on how a hairy BH may be anchored on the non-linearities of general relativity. 

As we have emphasized in this paper, these hairy MPBHs are disconnected, at the level of global properties, from the vacuum MPBHs, due to the mass (as well as other global quantities) gap\footnote{A mass gap for HBHs in $(2+1)$-dimensional AdS spacetime was pointed out in 
\cite{Stotyn:2012ap,Stotyn:2013spa}.}. 
But we would like to remark that, at the level of \textit{horizon} properties, they are actually connected to the vacuum MPBHs, as it is manifest in both Figures \ref{metric_limit} and \ref{areatemperature}.

Some preliminary results show that the solutions we have reported are not special to the case of MPBHs with symmetry enhancement. They also exist for arbitrary values of the angular momentum parameters and also in other dimensions. In the latter case there is, however, a qualitative difference. The finite mass gap we observed between the vacuum MPBHs and the marginally bound hairy MPBHs is also present in the horizonless limit of these solutions. Indeed, a mass gap exists between Minkowski spacetime and the marginally bound limit of boson stars~\cite{Hartmann:2010pm}. For $d>5$ boson stars have an \textit{infinite} mass gap. As such, we expect the hairy MPBHs in $d>5$ to have an infinite mass gap.

Finally, one may interpret the results in this work as showing that there are non-linear scalar clouds around MPBHs. This raises the question if non-linearities can also produce qualitatively new types of scalar clouds around Kerr BHs. We hope to report soon on this issue.

\vspace{0.5cm} 
\noindent
{\bf\large Acknowledgements}\\ 
The work of Y.B. was supported in part by an ARC contract AUWB-2010/15-UMONS-1. 
C.H and E.R. gratefully acknowledge support from the FCT-IF programme.  
The work in this paper is also supported by the grants PTDC/FIS/116625/2010 and  NRHEP--295189-FP7-PEOPLE-2011-IRSES.


\bibliographystyle{h-physrev4}
\bibliography{hbhs}
 
 \end{document}